\documentclass[manuscript]{aastex}

\usepackage{rotating}
\usepackage{tabularx}
\usepackage{amsmath}

\shorttitle{Properties of the July 23 CME-Driven Shock}
\shortauthors{Riley et al.}

\begin{document}


\title{Properties of the Fast Forward Shock Driven by the July 23 2012 Extreme Coronal Mass Ejection}


\author{Pete Riley, Ronald M. Caplan}
\affil{Predictive Science, 9990 Mesa Rim Road, Suite 170, San Diego, CA 92121, USA.}
\email{pete@predsci.com and caplanr@predsci.com}

\author{Joe Giacalone}
\affil{Lunar and Planetary Laboratory, University of Arizona, Tucson, AZ}
\email{giacalone@lpl.arizona.edu}

\author{David Lario}
\affil{Applied Physics Laboratory, The Johns Hopkins University, Laurel, MD}
\email{david.lario@jhuapl.edu}

\and 

\author{Ying Liu}
\affil{State Key Laboratory of Space Weather, National Space Science Center, 
Chinese Academy of Sciences, 
Beijing 100190, China}
\email{liuxying@spaceweather.ac.cn}

\begin{abstract}
Late on July 23, 2012, the STEREO-A spacecraft encountered a fast forward shock driven by a coronal mass ejection launched from the Sun earlier that same day. The estimated travel time of the disturbance ($\sim 20$ hrs), together with the massive magnetic field strengths measured within the ejecta ($> 100$nT), made it one of the most extreme events observed during the space era. In this study, we examine the properties of the shock wave. Because of an instrument malfunction, plasma measurements during the interval surrounding the CME were limited, and our approach has been modified to capitalize on the available measurements and suitable proxies, where possible. We were able to infer the following properties. First, the shock normal was pointing predominantly in the radial direction (${\bf n} = 0.97 {\bf e}_r -0.09 {\bf e}_t -0.23 {\bf e}_n$). Second, the angle between ${\bf n}$ and the upstream magnetic field, $\theta_{Bn}$, was estimated to be $\approx 34^{\circ}$, making the shock ``quasi-parallel,'' and supporting the idea of an earlier ``preconditioning'' ICME. Third, the shock speed was estimated to be $\approx 3300$ km s$^{-1}$.  Fourth, the sonic Mach number, $M_s$, for this shock was $\sim 28$. We support these results with an idealized numerical simulation of the ICME. Finally, we estimated the change in ram pressure upstream of the shock to be $\sim 5$ times larger than the pressure from the energetic particles, suggesting that this was not a  standard ``steady-state'' cosmic-ray modified shock (CRMS). Instead  it might represent an early, transient phase in the evolution of the CRMS. 
\end{abstract}


\keywords{Interplanetary Shocks; Extreme CMEs; Space Weather; Solar Wind; MHD Simulations}



\section{Introduction}
\label{Introduction} 
 
Coronal mass ejections (CMEs) are the spectacular result of the explosive release of energy stored in the magnetic field of the solar corona. While most CMEs propagate away from the Sun at speeds comparable to that of the ambient solar wind, a few have properties that could wreak havoc should they intercept Earth's orbital trajectory \citep{baker13a}. These so-called ``extreme events,'' whose likelihood of occurrence has been shown to follow a power law distribution \citep[e.g.][]{riley12a} and hence occur more frequently that might otherwise be anticipated, are still rare events. By definition, only a handful have been witnessed in modern times, the most famous of which is perhaps the Carrington event of 1859 \citep{carrington59a}.  On July 23, 2012, however, one of the two STEREO spacecraft measured an event of at least comparable strength. Thus, detailed studies of this event are crucial for improving our understanding of: (1) what conditions are required to these events to develop; (2) what physical processes produce such events; and (3) what properties can be anticipated.  

On July 23 2012, NASA's STEREO-A spacecraft, which was located at 0.96 AU and $121^{\circ}$ ahead of the Earth, measured what is undoubtedly the most extreme ICME observed during the space era. The structure, which was described in detail by \citet{russell13a}, consisted of a fast forward shock driven by what appeared to be a pair of abutting magnetic clouds, whose peak field strength exceeded 100 nT. EUVI observations by the SECCHI instrument suggested that a CME was launched at 0208 UT, while {\it in-situ} measurements suggest that the leading edge of the cloud arrived at 2255 UT, preceded by the shock at 2055 UT \citep{russell13a}. Thus, it took 20.78 hours for the CME to travel from the Sun to 1 AU, implying an average transit speed of $\sim 2,000$ km s$^{-1}$. 

At the Sun, the two prominence eruptions, separated by approximately 10-15 minutes likely produced the compound ICME at 1 AU \citep{liu14a}. Moreover, the observations by COR 2 onboard STEREO-B further suggest that the CME-driven shock was likely already well developed low in the corona. Time-elongation maps show two adjacent tracks supporting the idea of two CMEs being launched closely in time, and merging by the time they reached COR2's field of view \citep{liu14a}.  Estimates for the initial speed ranged from $2500 \pm 500$ km s$^{-1}$ \citep{baker13a} to 3,050 km s$^{-1}$ \citep{liu14a}.  

The July 23, 2012 CME and its interplanetary counterpart have been discussed in detail in a number of  other publications. \citet{liu14a} and \citet{temmer15a}, in particular, described the available remote solar observations made by both STEREO-A and STEREO-B and we do not discuss them further here. \citet{russell13a} considered the properties at, and upstream of the ICME proposing that the event represented a relatively unique example of a ``cosmic-ray-modified'' shock (CRMS). \citet{baker13a} considered the space weather implications of the event, addressing the likely consequences to the Earth's magnetosphere should an ICME like this have intercepted the Earth. 

In this study, we analyze the {\it in-situ} measurements by the STEREO-A spacecraft of the shock driven by the July 23 ICME. In particular, we focus on the properties of the shock driven by the extreme ICME as well as the energetic particle signatures associated with it, and we derive a set of simplified Rankine-Hugoniot relations that can be justified based on the limited availability of {\it in-situ} measurements. Additionally, we investigate the interval upstream of the shock to determine whether or not the shock conditions were modified in any significant way by the presence of energetic particles. To support our inferences, we have undertaken a suite of illustrative MHD simulations of pulse-like ejecta, which capture the basic features of the observations, including the properties of the shock. In Section~\ref{obs} we review the available observations of the shock. In Section~\ref{analysis-shock}, we infer the basic properties of the shock, including its orientation, the angle between the shock normal and the upstream magnetic field, and the shock speed. In Section~\ref{analysis-pressure} we focus on the various pressure terms associated with the region upstream of the shock to investigate whether this was a CRMS. Next, in Section~\ref{mhd} we develop a simple MHD model of the ejecta propagating from the upper corona to 1 AU, which captures the basic features of the observations. Finally, in Section~\ref{discussion} we summarize the main points of this study, discuss some of the limitations, assumptions, and caveats brought about by the incomplete observations and the idealized nature of the simulations. We also suggest areas that might be pursued to further our understanding of this unique event. 

\section{{\it In-Situ} Measurements from STEREO-A}
\label{obs}

In this section, we describe the main features of the available plasma, magnetic field, and energetic particle measurements. 

\subsection{Plasma Measurements}

The Plasma and Suprathermal Ion Composition (PLASTIC) investigation \citep{galvin08a} onboard STEREO-A produced the only available, but incomplete estimates of the plasma conditions during and surrounding the event. Initial estimates from the realtime Beacon feed were later shown to be incorrect \citep{baker13a}, with the maximum speed estimated to be $<1300$ km s$^{-1}$, and the complete absence of a shock front. Subsequently, more careful analysis of the PLASTIC data was only able to recover the bulk solar wind speed, not the plasma density, temperature, or transverse velocity components, either within, or upstream of the disturbance. 

To circumvent some of these issues, \citet{liu14a} used $>45$ eV electron density measurements as a proxy for the solar wind plasma density by multiplying each value by a factor of five. Similarly, they derived a plasma temperature based on the solar wind speed. For our purposes, the latter parameter probably cannot be used as a reliable estimate for plasma temperature variations during the event, and, at best, can be used as a proxy for the background value of the proton temperature upstream of the shock. The electron-derived density is more promising; however, we must be cautious of any quantitative estimates of shock parameters based on the inferred shock jump. 

The bottom panel of Figure~\ref{BBBBv} shows bulk solar wind speed as estimated from both proton fits (solid black line) and $O^{7+}$ and $O^{6+}$ fits (blue circles). Unfortunately, no directional information for the solar velocity has so far been recovered from the raw data. The vertical dashed lines mark boundaries associated with the event as identified by \citet{russell13a}. We note that each of the identified boundaries appear to be co-located with structure in the solar wind speed (1) a small jump in speed (boundary 1); (2) the start of a gradual rise in speed (boundary 2); (3) a sharp (at least within the resolution of the measurements) jump in speed corresponding to the location of a fast-mode shock \citep{russell13a} (boundary 3); and (4) a modest drop in speed (boundary 4). 

\subsection{Magnetic Field Measurements}

Measurements of the {\it in-situ} magnetic field were made by the IMPACT instrument \citep{luhmann08a}. The unit normal vectors of the field as well as the field magnitude are summarized in Figure~\ref{BBBBv}. Based on the smooth rotations of the magnetic field at the beginning of, and during a large interval during July 24, both \citet{russell13a} and \citet{liu14a} inferred the presence of the two ICMEs shown by the yellow shading. Although there is space between CME 1 and 2, the field strength profile suggests that the leading boundary of CME 2 may be advanced earlier in time, even to the point that no plasma separates the two structures. We note also that this boundary corresponds with the modest dip and rise in solar wind speed, which could be interpreted as a reverse shock, formed from the expansion of the first CME. 

The boundaries labeled 1 through 4 are clearly seen in the magnetic field measurements. Boundary 1 corresponds to a small rise in field strength, and a large change in the orientation of the interplanetary field, boundary 2, is seen as a drop in field strength, boundary 3 corresponds to the shock, while boundary 4 marks the leading edge of the complex ejecta. 

\subsection{Energetic Particle Measurements}

STEREO-A has three energetic ion detectors on board: SEPT, LET, and HET, each of which measures a progressively higher band of energies. Figure~\ref{ep-ts} summarizes these measurements during the time of transit of the CME and its associated disturbance. The individual traces show the measured intensity in a particular band, and each of the three thick curves shows the sum of the intensities over all energies measured by that instrument. We note following points, some of which were also made by \citet{russell13a}. First, the particle intensities increase dramatically shortly after the CME was launched at the Sun (02:08 UT). Second, the intensities flatten out between boundaries 1 and 2. Third, the intensities increase sharply between boundaries 2 and 3. Fourth, between boundaries 3 and 4, the intensities either decrease (HET), plateau (LET), or modestly increase (SEPT). Following boundary 4, the profiles all generally increase over several hours, then decrease, coincident with the inferred location of CME 1. Fifth, all intensities once again increase through CME 1. Sixth, beyond the trailing edge of CME 1, the intensities all decay to lower levels, flattening out by, or shortly after the trailing edge of CME 2. 

\section{Analysis of the Fast Forward Shock}
\label{analysis-shock}

As noted earlier, the fast forward shock was only partially observed by STEREO-A. Specifically, in terms of thermal plasma parameters, only the bulk solar wind velocity was captured and at relatively low resolution. We inferred a proton density across the shock by taking the measured electron density and multiplying it by a factor of five \citep{liu14a}. Magnetic field measurements, on the other hand were continuous throughout the interval with a cadence of 1 second (Figure~\ref{BBBBv}). 

Figure~\ref{ts-zoom} summarizes the three components of the magnetic field, the magnetic field strength, the `proton' number density, and the bulk solar wind speed. The magnetic field vectors have been smoothed and interpolated onto the resolution of the density (15-20 sec) to facilitate a more accurate calculation of the shock parameters. The yellow regions indicate the upstream (left) and downstream (right) intervals used in our shock analysis. The position of the shock is given by the dotted vertical line marked `3', consistent with the labeling by \citet{russell13a}. These intervals were chosen to best capture the asymptotic states on either side of the shock. Because of the low resolution of the plasma velocity data, the upstream window was moved earlier in time. It is worth noting that there does not appear to be any indication of self-excited magnetic fluctuations upstream of the shock: the components appear to be relatively smooth during the time period approaching boundary 3 (See also Figure~\ref{BBBBv}). 

There are a number of techniques for computing the properties of fast-mode shocks in the solar wind \citep[e.g.][]{riley96a}. However, given the availability and higher resolution of the magnetic field vectors, it makes sense to emphasize them in our analysis. Magnetic coplanarity, in particular, is a technique for estimating the orientation of the shock normal, relying on the fact that the  change in the magnetic field direction lies in the plane of the shock: 

\begin{equation}
\Delta {\bf B} \cdot {\bf n} = 0
\end{equation}
where $\Delta {\bf B} =  {\bf B}_2 - {\bf B}_1$. Thus, the unit normal can be written: 

\begin{equation}
{\bf n}_{mc} = \pm \frac{({\bf B}_2 \times {\bf B}_1) \times \Delta {\bf B}}{|({\bf B}_2 \times {\bf B}_1) \times \Delta {\bf B})|}
\label{n_mc}
\end{equation}

The relationship between these parameters is illustrated in Figure~\ref{sh-geom-schematics}. It is worth noting that the coplanarity approach breaks down in the limit that the angle between the unit normal and the upstream magnetic field, $\theta_{Bn} \rightarrow 0^{\circ}$ or $\theta_{Bn} \rightarrow 90^{\circ}$. In the classic picture of a fast-mode forward shock, both the magnetic field and velocity vectors are deflected away from the shock normal across the shock. And, while the bulk solar wind speed increases across the shock in the rest frame of the spacecraft (figures~\ref{BBBBv} and \ref{ts-zoom}), when transformed to the rest frame of the shock, the plasma slows down (c). Panel (c) also emphasizes how, since we only have information about the bulk solar wind, we must make an assumption that the flow is essentially radial across the shock. This, as we show below, is a reasonable approximation since the outward normal to the shock front is essentially radial, as is the observed solar wind flow upstream of the event. 
 
Rather than computing the shock normal (Equation~{\ref{n_mc}) once for the average parameters in the upstream and downstream regions, we have found it more reliable to match each point upstream with each point downstream, display the shock normal orientations graphically looking for clustering, then compute the average polar and azimuthal angles, $\theta_n$ and $\phi_n$, in heliographic coordinates \citep{riley96a}. This also allows us to estimate, at least heuristically, the likely uncertainties in our estimate. These results are shown in Figure~\ref{theta-phi}. 
 
 The individual shock orientations in Figure~\ref{theta-phi} have been color-coded according to whether they are outward-normals or inward normals. By convention, we consider only the outward-normals (red). We note that most are well clustered close to the radial direction.  The best estimate for the unit shock normal was: ${\bf n} = 0.97 {\bf e}_r -0.09 {\bf e}_t -0.23 {\bf e}_n$, or in terms of angular components, $\theta_{best} \sim -13.2^{\circ} \pm 10.2^{\circ}$ and $\phi_{best} \sim -5.6^{\circ} \pm 17.8^{\circ}$. 
 
 With the shock normal direction determined, we can then compute the angle between the upstream magnetic field and the shock normal, which has significance for understanding the types of particle acceleration processes that might be at work for this specific shock \citep[e.g.][]{zank00a}. This so-called $\theta_{Bn}$ parameter is defined by: 
 
 \begin{equation}
 \cos (\theta_{Bn}) = \frac{{\bf B}_1 \cdot {\bf n}}{B_1}
 \end{equation}
 
 Again, rather than computing a single value for the average upstream-downstream windows, we follow \citet{esparza96a} and pair each upstream point with each downstream point and compute all possible $\theta_{Bn}$ values. The distribution of these is shown in Figure~\ref{thetaBn}. The mean, median, and modal values all cluster at $\sim 34^{\circ}$, in agreement with a Gaussian fit to the histogram, suggesting that the shock could be described as quasi-parallel. 
 
To estimate the speed of the shock requires some consideration, given the lack of reliable plasma measurements. We begin by writing the jump relations for oblique (i.e., not purely parallel or perpendicular) shocks \citep{priest14a}:

 \begin{equation}
 \rho_2 v_{2x} = \rho_1 v_{1x}
\label{rh1}
 \end{equation}

 \begin{equation}
\rho_2 v_{2x} v_{2y} - \frac{B_{2x} B_{2y}}{ \mu} = \rho_1 v_{1x} v_{1y} - \frac{B_{1x} B_{1y}}{\mu}
 \end{equation}

\begin{multline}
(p_2 + \frac{B_2^2}{2 \mu}) v_{2x} - \frac{B_{2x}}{\mu} ({\bf B}_2 \cdot {\bf v}_2) + (\rho_2 e_2 + \frac{1}{2} \rho_2 v_2^2 + \frac{B_2^2}{2 \mu}) v_{2x} \\
= (p_1 + \frac{B_1^2}{2 \mu}) v_{1x} - \frac{B_{1x}}{\mu} ({\bf B}_1 \cdot {\bf v}_1) + (\rho_1 e_1 + \frac{1}{2} \rho_1 v_1^2 + \frac{B_1^2}{2 \mu}) v_{1x}
\end{multline}

 \begin{equation}
B_{2x} = B_{1x}
 \end{equation}

 \begin{equation}
v_{2x} B_{2y} - v_{2y} B_{2x} = v_{1x} B_{1y} - v_{1y} B_{1x}
\label{rh2}
 \end{equation}
where $\rho$, $v$, $B$, $p$, and $e$ have their usual meanings of density, speed, field strength, thermal pressure, and energy density. The subscripts $x$ and $y$ refer to directions parallel and perpendicular to the shock normal, and the 1's and 2's refer to conditions upstream and downstream, respectively. We note also that only thermal particles are included in the pressure term, as the energetic particle contribution to the pressure is usually neglected.  
Generally, we could use these relationships to derive a comprehensive $\chi$-squared minimization technique to estimate the properties of the shock (\citep[e.g.][]{vinas86a,szabo94a}; however, because of the limited information concerning the plasma parameters, particularly, the transverse velocity components, temperature, and to a lesser extent density, as well as the fact that the Alfv\'en Mach number is so large, and hence the magnetic-field terms play a relatively small role, we should limit ourselves to applying the first of these relations, mass conservation, which, in the spacecraft frame of reference, can be rewritten: 

 \begin{equation}
\rho_2 (v_{2x}' - v_{sh}') = \rho_1 (v_{1x}' - v_{sh}') 
 \end{equation}

Rearranging, we can estimate the speed of the shock, in the spacecraft's frame of reference, to be: 

 \begin{equation}
v'_{sh} = \frac{\rho_2 v_{2x}' - \rho_1 v_{1x}'}{\rho_2 - \rho_1}
 \end{equation}

Or, more generally: 

 \begin{equation}
v_{sh}' = \frac{1}{N} \sum\limits_{i=1}^N \frac{\Delta [\rho_i {\bf v'}_i] \cdot {\bf n}}{\Delta [ \rho_i]},
\label{sheq1}
 \end{equation}
where $i$ runs across all measurements within the upstream and downstream windows. Thus, we can estimate the speed of the shock relative to the upstream solar wind speed to be:

 \begin{equation}
v_{sh}^* = v_{sh}' - {\bf v}_{1}' \cdot {\bf n}
 \end{equation}

It is instructive to inquire about the sensitivity of the shock speed with respect to the density measurements. Assuming for now that $v'_x \rightarrow v'$, we can write the speed of the shock in the spacecraft frame of references as: 

 \begin{equation}
v_{sh}' \sim \frac{\rho_2 v_2' - \rho_1 v_1'}{\rho_2 - \rho_1}
 \end{equation}

Defining: $v_2' = v_1' + \Delta v'$, we can rewrite this expressions as follows:

 \begin{equation}
v_{sh}' \sim \frac{\Delta v'}{1-\rho_1/\rho_2} + v_1'
 \end{equation}

Thus, for a strong shock: $ \frac{\rho_2}{\rho_1} \rightarrow 4$

\begin{equation}
v_{sh}' \rightarrow \frac{1}{3} (4 v_2' -  v_1')
\label{sheq2}
\end{equation}

%

Using Equation~(\ref{sheq1}), we can estimate the speed of the shock in the rest frame of the spacecraft based on the jump in density to obtain: $v_{sh}' = 3377$ km s$^{-1}$. As a rudimentary check, we can compare this number with the value we would estimate in the limit of a strong shock (Equation~(\ref{sheq2})), $v'_{sh} \rightarrow \frac{1}{3} (4 \times 2250 -  900) \sim 2700$ km s$^{-1}$. As an even more basic check, in the limit that the shock were perpendicular, Equation(~\ref{rh2}) implies that the ratio of the upstream to downstream perpendicular components of the magnetic field mimic the jump in density (Equation~(\ref{rh1})), thus, $v_{sh}' (B_{\perp}) = 2752$ km s$^{-1}$. These are, of course, gross oversimplifications. Nevertheless, they provide basic support for the estimated shock speed of 3377 km s$^{-1}$. 

Finally, we can estimate the Sonic and Alfv\'en Mach numbers for this event. The Sonic Mach number is given by:

\begin{equation}
M_s = \frac{v_1}{C_{s1}}
\end{equation}
where $v_1$ is the upstream solar wind speed in the shock's frame of reference and $C_{s1} = (\gamma P_1/\rho_1)^{\frac{1}{2}}$. Assuming $P_{th} = 2 n k_B T$, allows us to write $C_{s1} = (2 \gamma k_B T_1/m_p)^{\frac{1}{2}}$, where $k_B$ is the Boltzmann constant and $m_p$ is the mass of a proton. Setting $\gamma = 5/3$ and using the inferred - but likely unreliable - proton temperature upstream of the shock \citep{liu14a} of $T_p \sim 2 \times 10^5$ K results in $C_{s1} \sim 74$ km s$^{-1}$, and, hence, with $v_1 \sim 2100$ km s$^{-1}$ (see Figure~\ref{VswVsh}), a sonic Mach number, $M_s \sim 28$. Assuming a canonical value for the Alv\'en speed in the solar wind, $V_A \sim 100$ km s$^{-1}$, produces an estimate for the Alfv\'en Mach number, $M_A \sim 21$. 

\section{Analysis of the Ram, Thermal, Magnetic, and Energetic Particle Pressures}
\label{analysis-pressure}

We next turn our attention to various pressure terms related to the ICME and its associated disturbance. In particular, we want to understand whether, and/or to what extent, the July 23, 2012 event was a CRMS, as was proposed by \citet{russell13a}. 

\citet{russell13a} demonstrated that the pressure exerted by the energetic particles ($P_{ep}$) significantly exceeded that of the magnetic pressure ($P_{mag}$). However, \citet{terasawa99a} suggested that, at least within the limitations of a steady-state model, the determining quantity to compare against was the change in the ram pressure of the gas, $P_{ram} = \rho v \Delta v$. Thus, we would like to know whether the total pressure, $P_{total}$ ($=P_{ep} + P_{th} + P_{mag}$) balances the change in ram pressure, $\Delta P_{ram}$. 

The thermal and magnetic pressures are computed as usual, with $P_{th} = 2 n_p k_B T_p$ and $P_{mag} = \frac{B^2}{2 \mu_0}$ \citep[e.g.][]{riley98a}. The pressure exerted by the energetic particles is estimated as follows: 

\begin{equation}
P_{ep} = 4 \pi m_p \sum\limits_{i=1}^N I v \Delta E
\end{equation}
where $v = \sqrt{\frac{2 E}{m_p}}$, $I$ is the measured intensity, $\Delta E$ is width of each energy bin, and $i$ runs from 1 to $N$, the total number of energy bins. 

Figure~\ref{Pcomp} compares the relative pressure terms, including the ram, magnetic, energetic particle, and gas pressures. it also breaks out the contributions within the three different instruments (LET, SEPT, and HET). We note that the ram pressure always dominates over the other pressure terms by as much as two orders of magnitude. However, and as pointed out by \citet{russell13a}, although the magnetic pressure term is generally much larger than the energetic particle pressure, this is not the case ahead of the shock front (boundary 3), and particularly in the region immediately ahead of it (between boundaries 2 and 3). There, the $P_{ep} >> P_{mag}$. 

However, following the suggestion by \citet{terasawa99a}, we should consider how the energetic particle pressure compares with the change in ram pressure. To do this, we must first transform the speed measurements from their spacecraft frame of reference to that of the shock frame. This is shown in Figure~\ref{VswVsh}. We note that, in passing across the shock (boundary 3), the solar wind flow slows from $>2000$ km s$^{-1}$ to $\sim 750$ kms$^{-1}$. Setting the change in the ram pressure to be zero at the location of boundary 2, we can then compute $\Delta P_{ram}$. This is shown and compared with the changes in the other magnetic pressure terms in Figure~\ref{dPPPP}. We make the following remarks. First, $P_{ep}$ dominates over both $P_{mag}$ and $P_{th}$. Second, the change in the ram pressure dominates over any of the other pressure terms, including the energetic particle pressure as well as the sum of them. Third, the variations in $P_{ep}$ and $P_{ram}$ are, however, in the opposite sense such that, at least to some extent, they offset one another. 

\section{MHD Modeling}
\label{mhd}

To confirm the reasonableness of our analysis, we also developed some simple numerical simulations mimicking the launch of a CME and its propagation through the inner heliosphere. The model is highly idealized, intended only to provide basic support for the inferences drawn from the observations. 

\subsection{Model Description}

To simulate the July 23, 2012 ICME we used the PLUTO astrophysical MHD code \citep{mignone07a}. The code is ideally suited for solving the MHD equations under conditions of high Mach number flows. The code is modern, modular, and user-friendly, allowing the user to easily modify the boundary conditions for custom runs. It also contains a number of sophisticated features that weren't necessary for the simulations we undertook, including Adaptive Mesh Refinement (AMR) and the ability to run on massively parallel architectures. The runs summarized here were performed on a desktop computer.  Given that only one component of the velocity was measured ($V_r$), we ran the code in a 1-D spherical coordinate system ($r$), first setting the inner radial boundary at 30 $R_S$ to some reasonable ambient values \citep[e.g.][]{riley97a,riley98a} and allowing the solution to achieve a steady-state equilibrium. We set the mean molecular weight, $\mu = 0.6$ to reflect a completely ionized plasma primarily consisting primarily of hydrogen with a small component of helium. For simplicity, we set the gravitational constant to zero, a reasonable approximation for simulations starting in the high corona. 
We then introduced a perturbation at the inner boundary to mimic the passage of a fast ICME through it. We varied the initial speed, density, and duration of the initially smoothly-varying ($\sin^2$ profile) pulse in an attempt to mimic the gross features of the ejecta and associated shock observed at 1 AU. Unlike most simulations of ICMEs in the inner heliosphere, since we are restricting the model to 1-D we are also able to specify a transverse component of the magnetic field within the ejecta, thus providing a more accurate contribution to the magnetic pressure of the ICME. Of course, we cannot model any toroidal field pattern, which, in reality was observed for this event.

\subsection{Model Results}
\label{results}

Figure~\ref{pluto} illustrates one of the simulation results. Panels (a) through (d) show time series of the speed, density, magnetic field strength, and temperature of the ICME. The start of the time axis is arbitrary in the sense that it is measured from the initiation of the simulation, which included a substantial period for the solution to reach equilibrium prior to the launch of the CME. Comparing with the observations, the speed profile matches relatively well, with an abrupt jump up to almost 2500 km s$^{-1}$ and subsequent decay. There is even the suggestion of a reverse wave in the observations during the early declining speed profile, which is quite prominent in the model. Since neither density or temperature could be recovered from the plasma instrument data, we can only speculate on whether these are good facsimiles of the observations. The magnetic field measurements appear reasonable at the leading portion of the event, but, since we do not include what we believe was a second ICME following the first one, we cannot address the structure within the trailing portion of the event. 

In Figure~\ref{pluto}(d) we show the speed of the solar wind and embedded ICME as a function of time (x-axis) and distance from the Sun (y-axis). This illustrates how an initially super-fast velocity pulse bifurcates into two distinct waves; a fast forward wave to the left, and a fast reverse wave further to the right. With increasing distance from the Sun, the two separate. We can also use this visualization to directly estimate the speed of the modeled fast-mode shock. The solid white line is drawn tangent to the shock front at 1 AU, and has a slope of 2,967 km s$^{-1}$, roughly consistent with the speeds estimated using the {\it in-situ} measurements above. 

\section{Discussion}
\label{discussion} 

In summary, our study suggests that the shock driven by the July 23/24, ICME complex was propagating approximately radially (${\bf n} = 0.97 {\bf e}_r -0.09 {\bf e}_t -0.23 {\bf e}_n$), with a speed of $ \sim 3300$ kms$^{-1}$ in the spacecraft frame of reference. The upstream speed was estimated to be $\sim 2100$ kms$^{-1}$, suggesting that the Mach number of the shock, $M_s \sim 28$. Finally, the angle between the upstream magnetic field and the shock normal was estimated to be $\sim 34^{\circ}$. Additionally, our idealized MHD simulations support the basic inferences from the observations. In particular, the shock was strong, with the ratio of downstream to upstream densities, $\rho_2/\rho_1 \sim 4$ and the speed of the shock estimated to be $\sim 3000$ kms$^{-1}$. 

Our analysis of the various pressure terms surrounding the shock suggest that the energetic particle pressure was not sufficiently large to offset the change in ram pressure. At most, it could account for 20\%. Thus, we find no evidence that this was an energetic particle-mediated shock, as suggested by \cite{russell13a}. It should be clarified, however, that our analysis does not preclude the possibility that the shock was not modified by the presence of the energetic particles. Indeed, it could be argued that the boundary marked `2' in Figures~\ref{BBBBv} and \ref{ep-ts}, together with the interval between `2' and `3' are evidence that some form of shock mediation took place. However, there are a number of other explanations for these variations, not least of which is that the solar wind is not generally observed to be in pressure balance along a radial trace \citep[e.g.][]{mccomas96a}. Importantly, \citet{terasawa99a} reasoned that, at least within the context of a steady-state model, the change in ram pressure, $\rho v \delta v$, should be balanced by the increase in cosmic ray pressure. Instead, we found here, that although the changes did attempt to offset one another, the pressure from the energetic particles could only account for 20\% of the change in the ram pressure. One possible interpretation is that the steady-state model does not apply here and that what we observed was an early, transient phase of the modification of the upstream properties of the shock by the energetic particles. If we were able to observe it evolve, we may have seen the pressure exerted by the energetic particles continue to increase until, at some point, it would have matched the variation in ram pressure. This would likely have resulted in more significant substantial modifications to the magnetic field and thermal plasma properties. 

We must also recognize that our computation of the energetic particle pressure is likely an underestimate. The instruments onboard the STEREO spacecraft, while capturing a large dynamic range of energies cannot capture all of them. In particular, there may be a significant, and unmeasured component between the thermal peak of the plasma and the lowest-energy of the energetic particle detectors ($~\sim 20-30$ keV). We do not believe, however, that such a component could represent an additional 400\%, which is what would be required to balance the offset in ram pressure. 

It is worth re-emphasizing that the speed of the CME-driven shock was likely in excess of 3,000 km s$^{-1}$. Several studies reporting on this event have erroneously stated that the speed of the shock was 2250 km s$^{-1}$ \citep[e.g.][]{temmer15a}. This is the speed of the bulk plasma flow at the time the shock passed the spacecraft, not the speed of the shock itself. In fact, from simple shock jump condition analysis, the shock speed must be faster than the peak speed of the disturbance, otherwise flow upstream of the shock front could not flow into the shock and be decelerated. 

Our results are consistent with the suggestion that an earlier ICME preconditioned the solar wind environment \citep{liu14a} to create circumstances that promoted the extreme character of the event. Additionally, our shock analysis suggests that $\theta_{Bn}$, the angle between the shock normal and the upstream magnetic field, was quasi-parallel. Since we determined that the shock front was propagating roughly in the radial direction (tiled modestly toward the equator and slightly eastward), this suggests that the pitch of the spiral field was less than the nominal $\approx 45^{\circ}$ based on the Parker estimate for 400 km s{$^{-1}$. In general, fast ICMEs propagating through the solar wind tend to produce under-wound magnetic fields lines, sometimes, even radial fields \citep[e.g.][]{riley07e}. 

In closing, the July 23, 2012 ICME was an ``extreme'' event by any standard. In particular, the speed and strength of the shock driven by it would have initiated a strong magnetospheric response, which would have been further amplified by the large, complex ICME structure that followed \citep{baker13a}.  

\begin{acknowledgments}
PR gratefully acknowledges the support of NASA (LWS Workshop on Extreme Space Weather Events) and NSF (Frontiers in Earth System Dynamics (FESD) program). DL acknowledges the support from NASA under grants NNX11A083G and NNX15AD03G.
\end{acknowledgments}

\clearpage



\begin{figure}[tb]
 \centering
 \resizebox{0.75\textwidth}{!}{
 \includegraphics{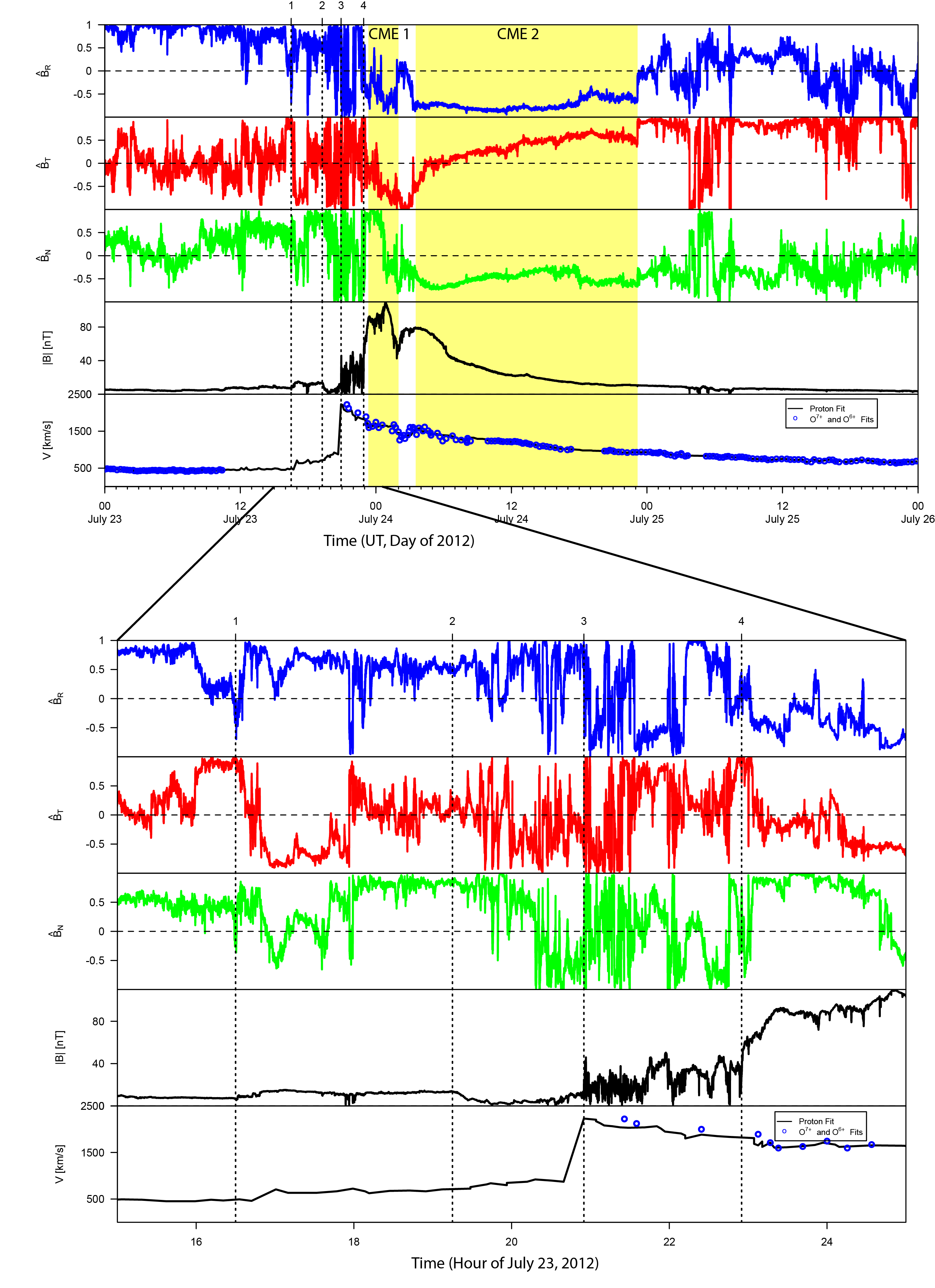}}
 \caption{
STEREO-A {\it In-situ} measurements of the unit-normal components of the interplanetary magnetic field and bulk solar wind speed as a function of time from July 23 through July 25, 2012. Two estimates of the speed are shown: proton fits (solid black line) and $O^{7+}$ and $O^{6+}$ fits (blue circles). The four dashed vertical lines are the boundaries identified by \citet{russell13a}, while the two yellow boxes indicate the inferred location of the two ICMEs making up the complex ejecta \citep{liu14a}.  The lower panel provides more detailed view of the time interval containing the boundaries 1-4.
} \label{BBBBv}
\end{figure}
 
\begin{figure}[tb]
 \centering
 \resizebox{0.85\textwidth}{!}{
 \includegraphics{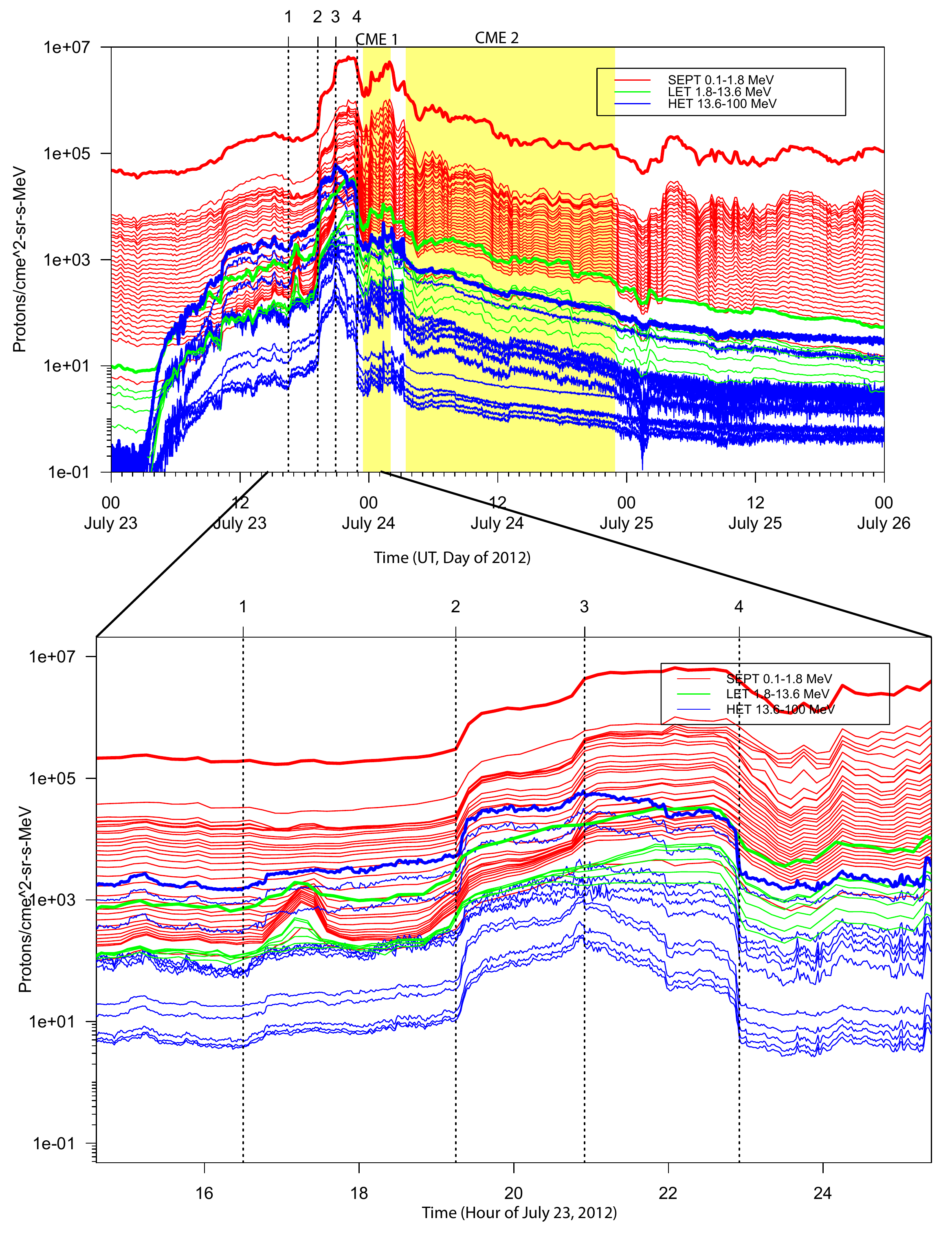}}
 \caption{
{\it In-situ} measurements of energetic proton intensities prior to, and throughout the ICME interval. Red curves are from the SEPT instrument, which measures the energy range 0.1 - 1.8 MeV, green curves are from the LET instrument, which measures from 1.8 to 13.6 MeV, and blue curves are from the HET instrument, measuring from 13.6 to 100 MeV.   
} \label{ep-ts}
\end{figure}
 

\begin{figure}[tb]
 \centering
 \resizebox{0.95\textwidth}{!}{
 \includegraphics{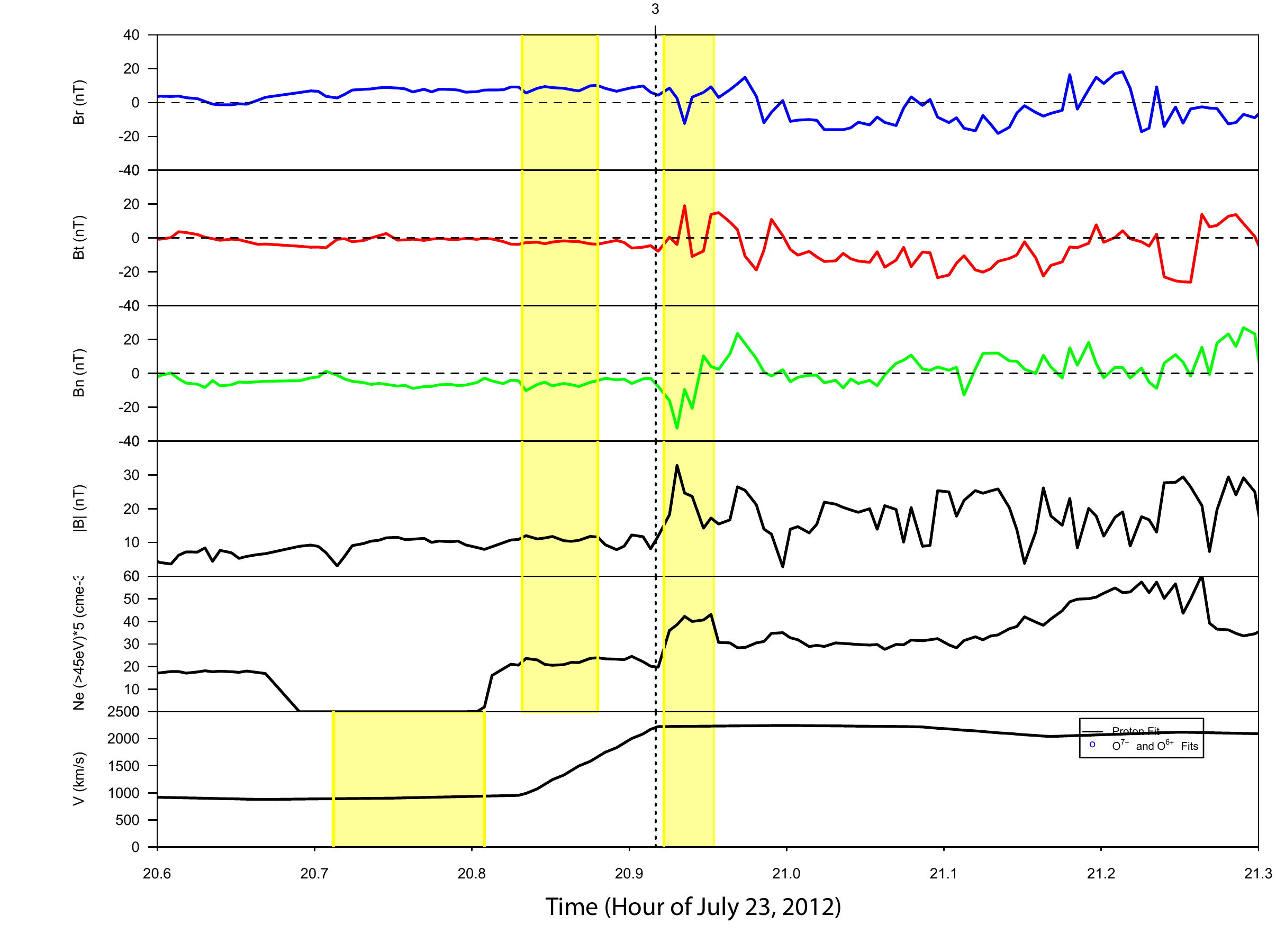}}
 \caption{
Magnetic field vectors and speed across the shock front. The upstream and downstream intervals used in the analysis are shown by the yellow shaded regions. The proxy plasma density is constructed by multiplying the electron energy ($> 45$ eV) by five. 
} \label{ts-zoom}
\end{figure}
 
 \begin{figure}[tb]
 \centering
 \resizebox{0.95\textwidth}{!}{
 \includegraphics{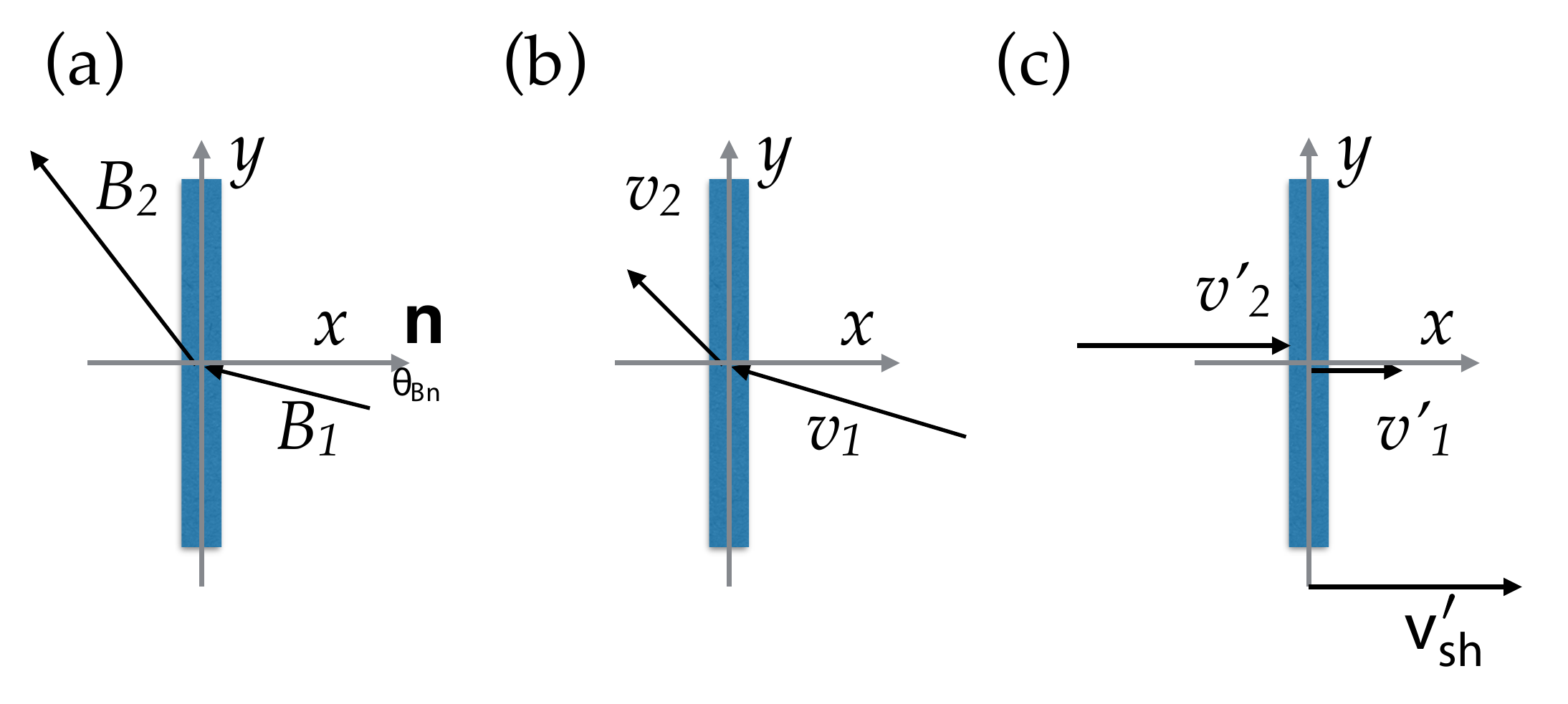}}
 \caption{
Illustration of the jump in parameters across a fast-forward shock: (a) the magnetic field vector; and (b) the velocity vector in the rest frame of the shock. (c) The change in speed in the in the spacecraft's frame of reference, assuming that ${\bf v}' \sim v'_r$. The orientation of the shock normal, {\bf n}, relative to the shock front is also shown as well as the angle between the shock normal and the upstream magnetic field. Upstream parameters are denoted by the subscript `1' while downstream parameters are denoted by `2'. 
} \label{sh-geom-schematics}
\end{figure}
 
\begin{figure}[tb]
 \centering
 \resizebox{0.95\textwidth}{!}{
 \includegraphics{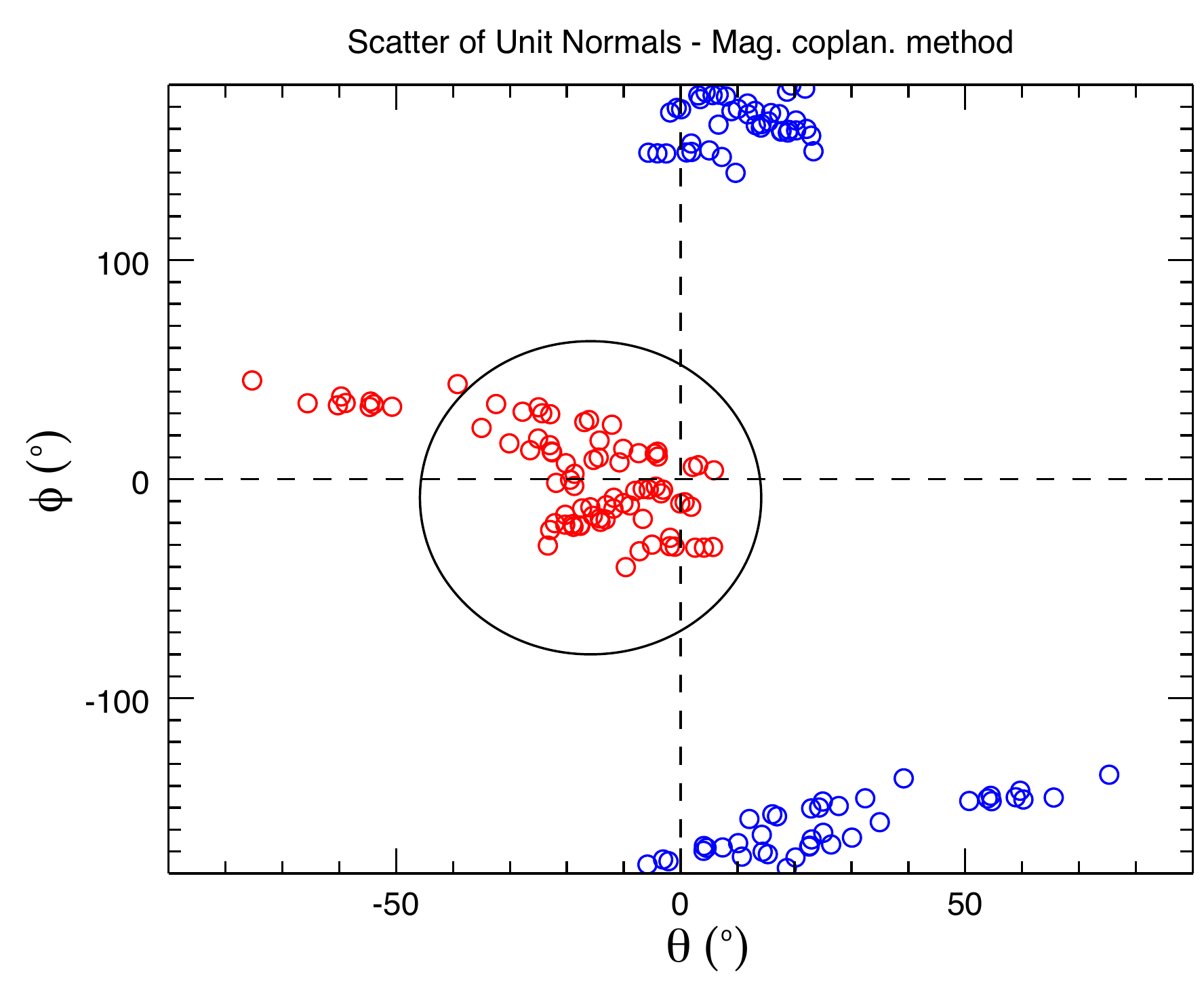}}
 \caption{
The location of the forward (red) and reverse (blue) shock normal angles as determined using magnetic coplanarity by matching all data points within each upstream and downstream region with one another. Theta ($\theta$) is the meridional angle, positive from the north pole southward. Phi ($\phi$) is the azimuthal tilt of the shock front, positive being in the westward direction. The outward radial direction is at (0,0). The circle defines the cluster of points used to construct the best estimate and uncertainty.   
} \label{theta-phi}
\end{figure}

\begin{figure}[tb]
 \centering
 \resizebox{0.95\textwidth}{!}{
 \includegraphics{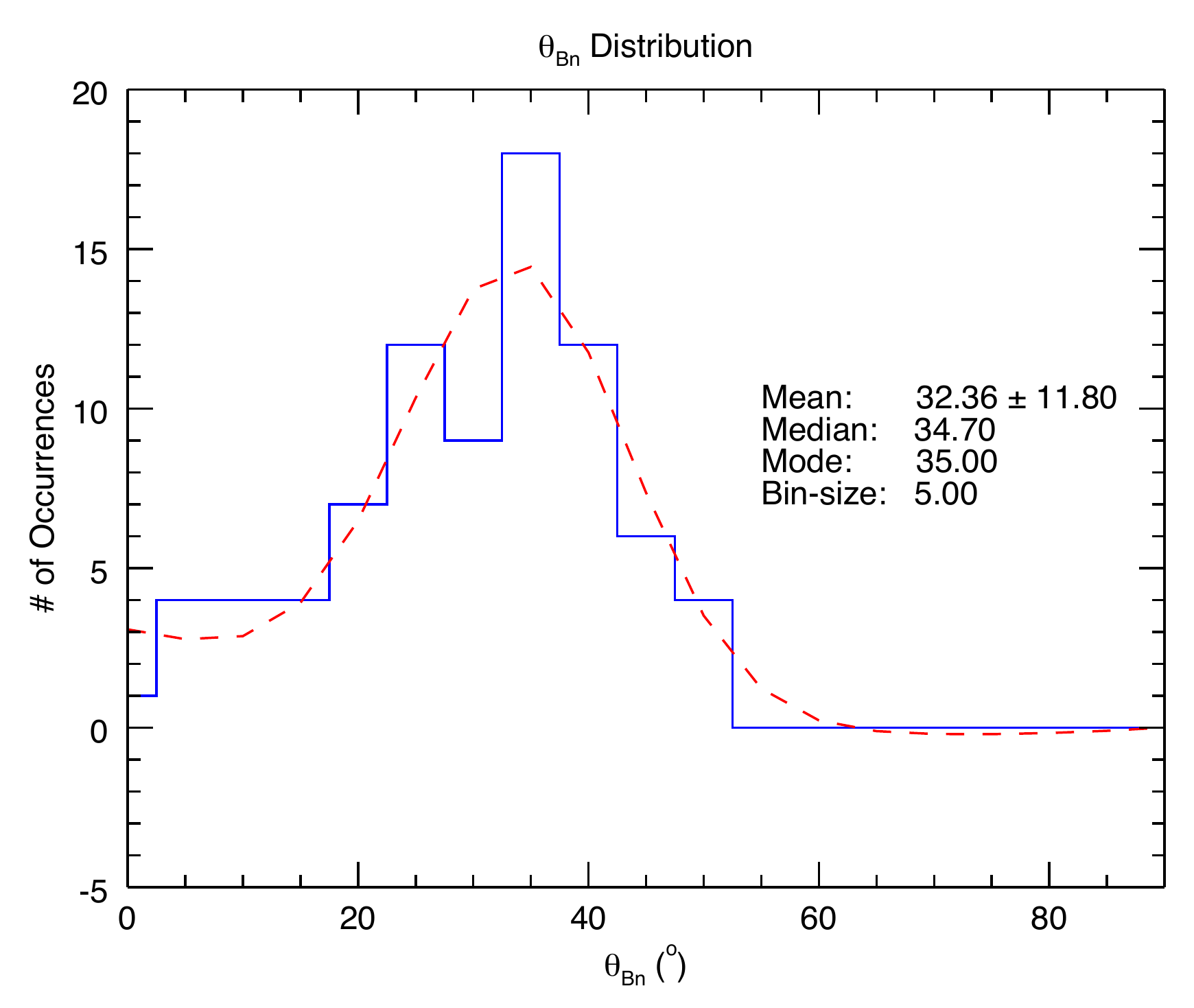}}
 \caption{
Distribution of $\theta_{Bn}$ angles for all pairings of the data points in the windows marked in Figure~\ref{ts-zoom}.   The mean, median, and modal values are also shown, together with a Gaussian fit to the histogram.
} \label{thetaBn}
\end{figure}

\begin{figure}[tb]
 \centering
 \resizebox{0.75\textwidth}{!}{
 \includegraphics{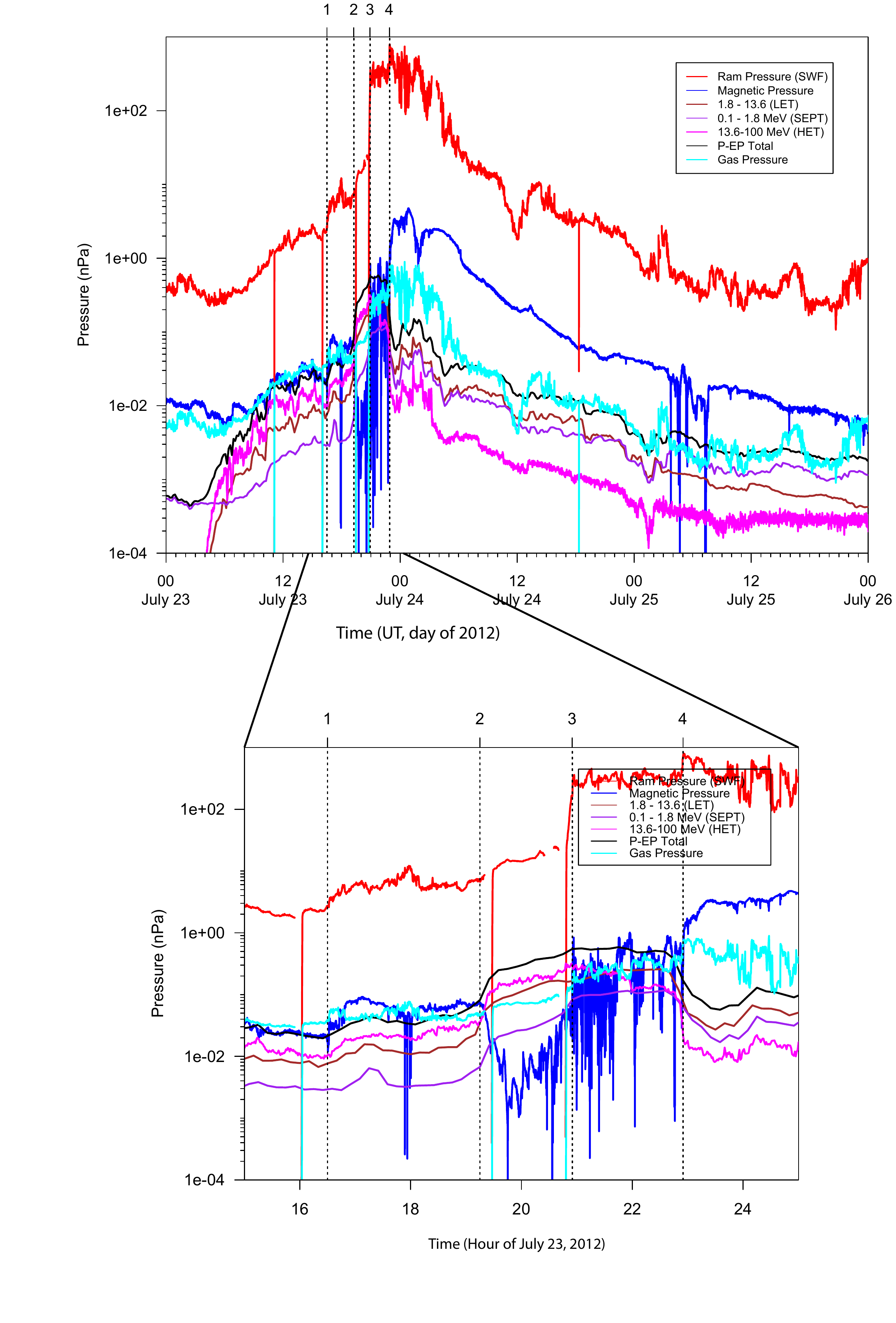}}
 \caption{
Comparison of various plasma and magnetic field pressures: Ram pressure (red); magnetic pressure (blue); three components of energetic particle pressure (brown, purple, and magenta); total energetic particle pressure (black); and gas pressure (aqua). The four vertical dashed lines (1-4) are the same as those in Figure~\ref{BBBBv}. 
} \label{Pcomp}
\end{figure}

\begin{figure}[tb]
 \centering
 \resizebox{0.95\textwidth}{!}{
 \includegraphics{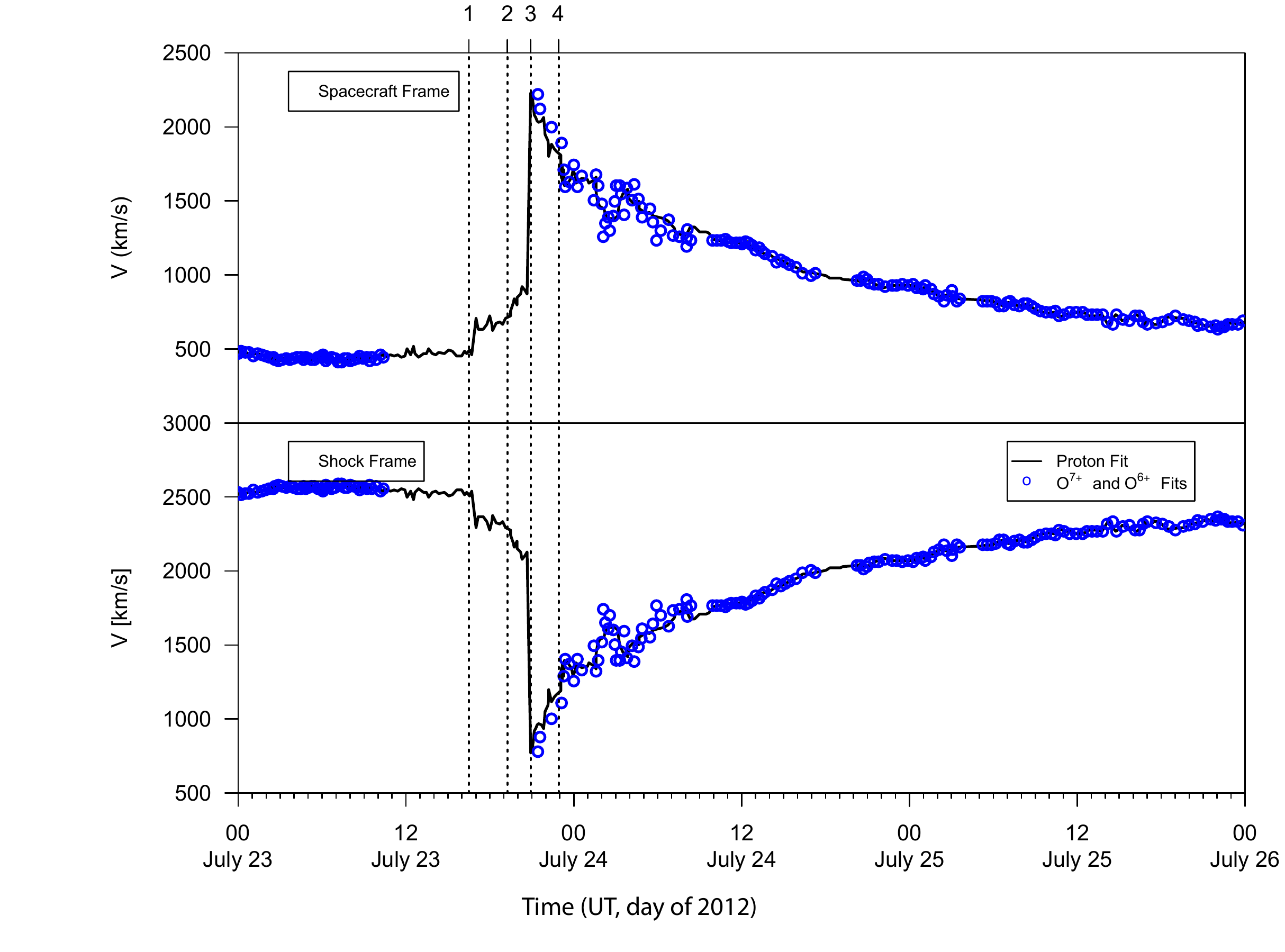}}
 \caption{
Time series of solar wind speed in (top) spacecraft frame and (bottom) shock frame. The solid black lines are based on proton fits while the blue circles are based on fits to $O^{7+}$ and $O^{6+}$. 
} \label{VswVsh}
\end{figure}

\begin{figure}[tb]
 \centering
 \resizebox{0.95\textwidth}{!}{
 \includegraphics{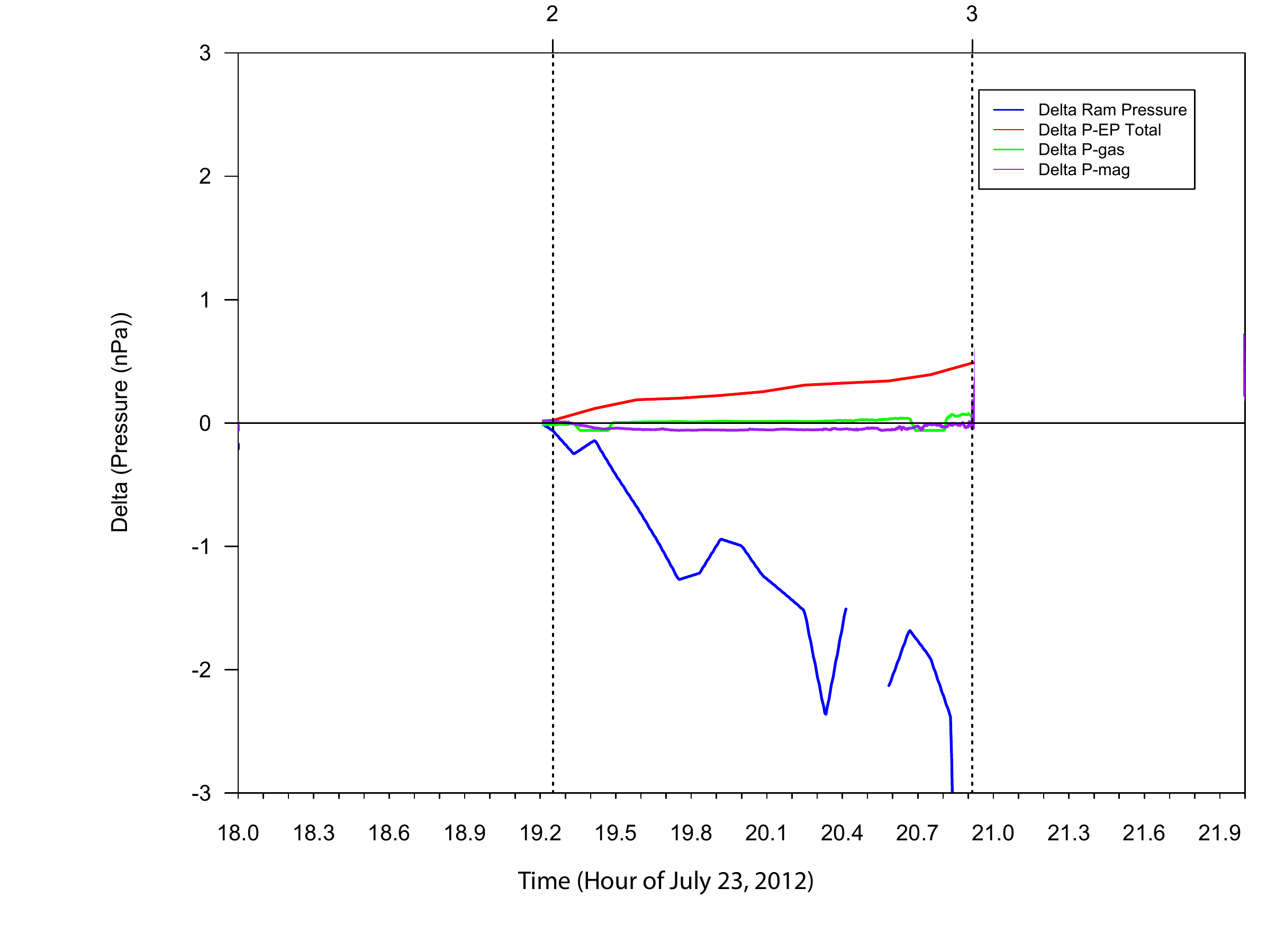}}
 \caption{
Time series of the change in ram (blue), energetic particle (red), gas (green), and magnetic (purple) pressures for the region bounded between boundaries 2 and 3, i.e., immediately upstream of the shock. 
} \label{dPPPP}
\end{figure}

\begin{figure}[tb]
 \centering
 \resizebox{0.95\textwidth}{!}{
 \includegraphics{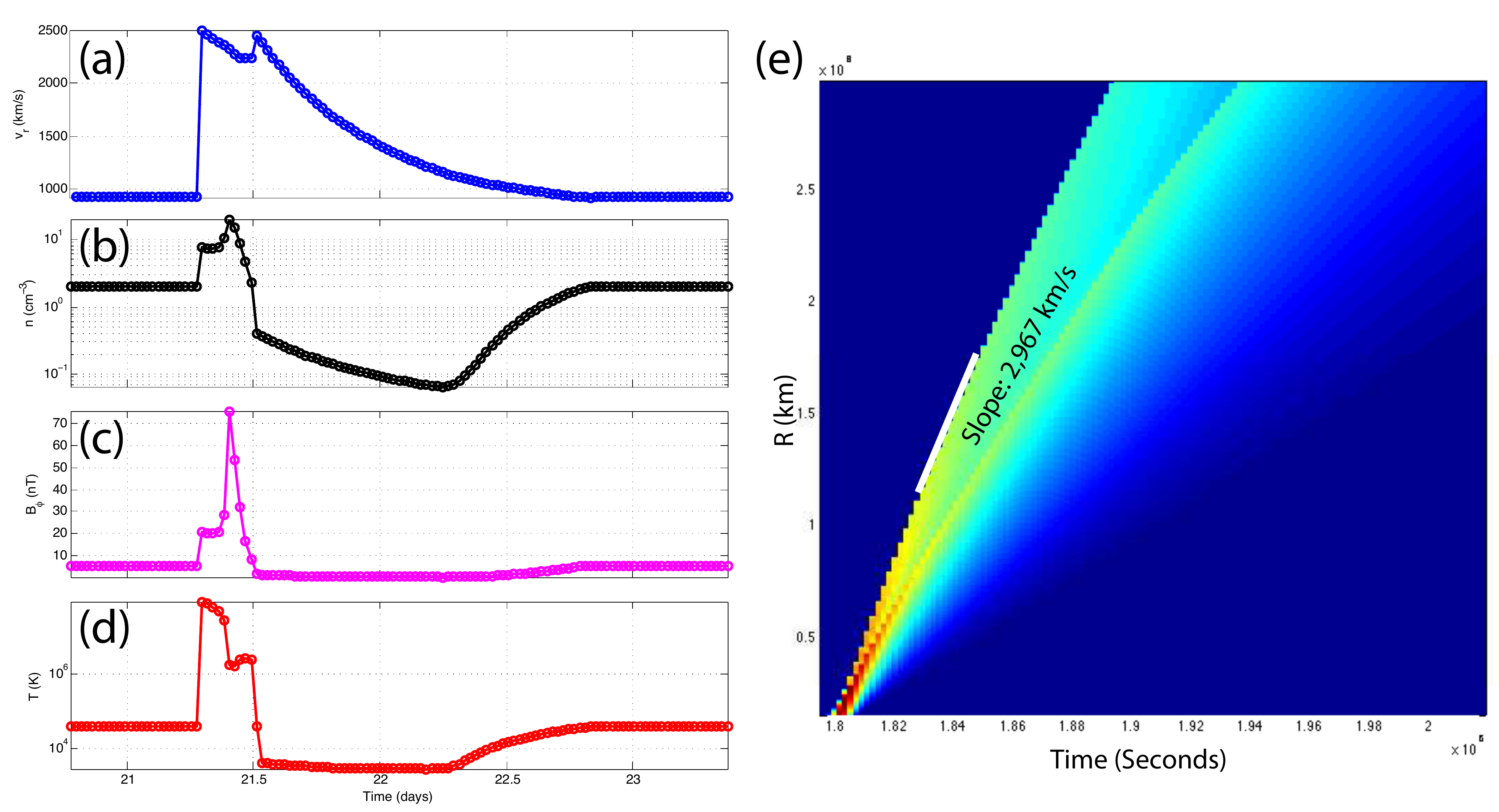}}
 \caption{
Results from an idealized 1-D MHD simulation using the Pluto code. (a) - (d) time series of velocity ($v$), density ($n$), transverse magnetic field ($B$), and temperature ($T$) at 1 AU. (e) Velocity map showing the evolution of speed as a function of time (x-axis) and distance from the Sun (y-axis). The white line is a tangent to the location of the shock at $\sim 1$ AU. 
} \label{pluto}
\end{figure}

\newpage







\clearpage




\clearpage




\end{document}